\documentclass[journal]{IEEEtran}

\ifCLASSINFOpdf
\usepackage[pdftex]{graphicx}
\else
\usepackage[dvips]{graphicx}
\fi
\usepackage{amsmath}
\usepackage{amsmath} 
\usepackage{amsfonts}
\usepackage{multicol}
\usepackage{enumerate}
\begin{document}
%
\title{SVD-Aided Multi-Beam Directional Modulation Scheme Based on Frequency Diverse Array}

\author{Qian~Cheng,~\IEEEmembership{Student Member, IEEE},
        Vincent~Fusco,~\IEEEmembership{Fellow, IEEE},
        Jiang~Zhu,~\IEEEmembership{Member, IEEE},
        Shilian~Wang,~\IEEEmembership{Member, IEEE},
        and~Chao~Gu
\thanks{This work was supported by a scholarship (No. 201803170247) from China Scholarship Council (CSC). \emph{(Corresponding author: Shilian Wang.)}}
\thanks{Q. Cheng is with the College of Electronic Science, National University of Defense Technology, Changsha 410073, China, and also with the ECIT Institute, Queen's University Belfast, Belfast BT3 9DT, UK. (e-mail: chengqian14a@nudt.edu.cn)}
\thanks{J. Zhu and S. Wang are with the College of Electronic Science, National University of Defense Technology, Changsha 410073, China (e-mail: \{jiangzhu, wangsl\}@nudt.edu.cn).}
\thanks{V. Fusco and C. Gu are with the ECIT Institute, Queen's University Belfast, Belfast BT3 9DT, UK (e-mail: v.fusco@ecit.qub.ac.uk; chao.gu@qub.ac.uk).}
}

\maketitle

\begin{abstract}
With the assistance of singular value decomposition (SVD), a multi-beam directional modulation (DM) scheme based on symmetrical multi-carrier frequency diverse array (FDA) is proposed. The proposed DM scheme is capable of achieving range-angle dependent physical layer secure (PLS) transmissions in free space with much lower complexity than the conventional zero-forcing (ZF) method. Theoretical and simulated results about secrecy rate and complexity verify the improved computational efficiency and considerable memory savings despite a very small penalty of secrecy rate.

\end{abstract}


\IEEEpeerreviewmaketitle

\section{Introduction}

\IEEEPARstart{D}{irectional} modulation (DM) has increasingly become a promising technique capable of achieving physical layer secure (PLS) communications.

The work in \cite{Daly_DM} proposed an angle-dependent DM scheme based on phased arrays (PA) by optimizing the phase shifters at the radio frequency (RF) frontend. A low-complexity DM synthesis method was proposed in \cite{Ding_Vector}, which transferred the DM synthesis from RF frontend to baseband. Afterwards, multi-beam angle-dependent DM synthesis methods were researched in \cite{Ding_Orthogonal}-\cite{Xie_AN_ZF}. Specifically, the orthogonal vector synthesis approach was implemented in \cite{Ding_Orthogonal}, while \cite{Shu_Robust} proposed a robust synthesis method for multi-beam DM scheme with imperfect direction knowledge. The multipath nature of the channel was exploited in \cite{Hafez_SM_DM} to create a multi-user DM transmissions. Artificial noise (AN) aided multi-beam DM system was studied in \cite{Xie_AN_ZF} using zero-forcing (ZF) criterion.

The above-mentioned single- or multi-beam DM schemes can only achieve angle-dependent secure transmissions, the security of which will fail when the eavesdropper is located along the same direction as the desired receiver. To address this problem, the frequency diverse array (FDA) with non-linear frequency increments was introduced to synthesize both range-angle dependent DM in \cite{Wang_FDA_DM}. The work in \cite{Hu_Random_FDA_DM} proposed a random FDA-based range-angle dependent DM synthesis method, while the AN-aided FDA-DM communication over Nakagami-\textit{m} fading channels was studied in \cite{Ji_FDA_DM_Fading}. 

These FDA-DM schemes \cite{Wang_FDA_DM}-\cite{Ji_FDA_DM_Fading}   only considered single-beam DM transmissions for a single desired receiver. To achieve range-angle dependent multi-beam FDA-based DM transmissions, one possible approach \cite{Qiu_MB_FDA_DM} is to jointly optimizing the frequency increments, the beamforming vector, and the AN orthogonal matrix, which is too complicated to realize practically. The other approach is to extend the ZF-aided multi-beam PA-DM  synthesis \cite{Xie_AN_ZF} to multi-beam FDA-DM synthesis \cite{Xie_FDA_DM}. The ZF method, however, consumes a large amount of memory for the orthogonal matrix and the AN vector.

This paper aims to reduce the complexity of the range-angle dependent multi-beam DM system in free space by re-designing the AN vector and orthogonal matrix with the aid of singular value decomposition (SVD). Compared with the conventional ZF method, the proposed SVD method can improve computational efficiency and save much memory with a small loss of secrecy rate. This makes it possible to achieve secure range-angle dependent multi-beam DM transmissions with lower complexity and lower DC power consumption.

\begin{figure}
\centering
\includegraphics[angle=0,width=0.45\textwidth]{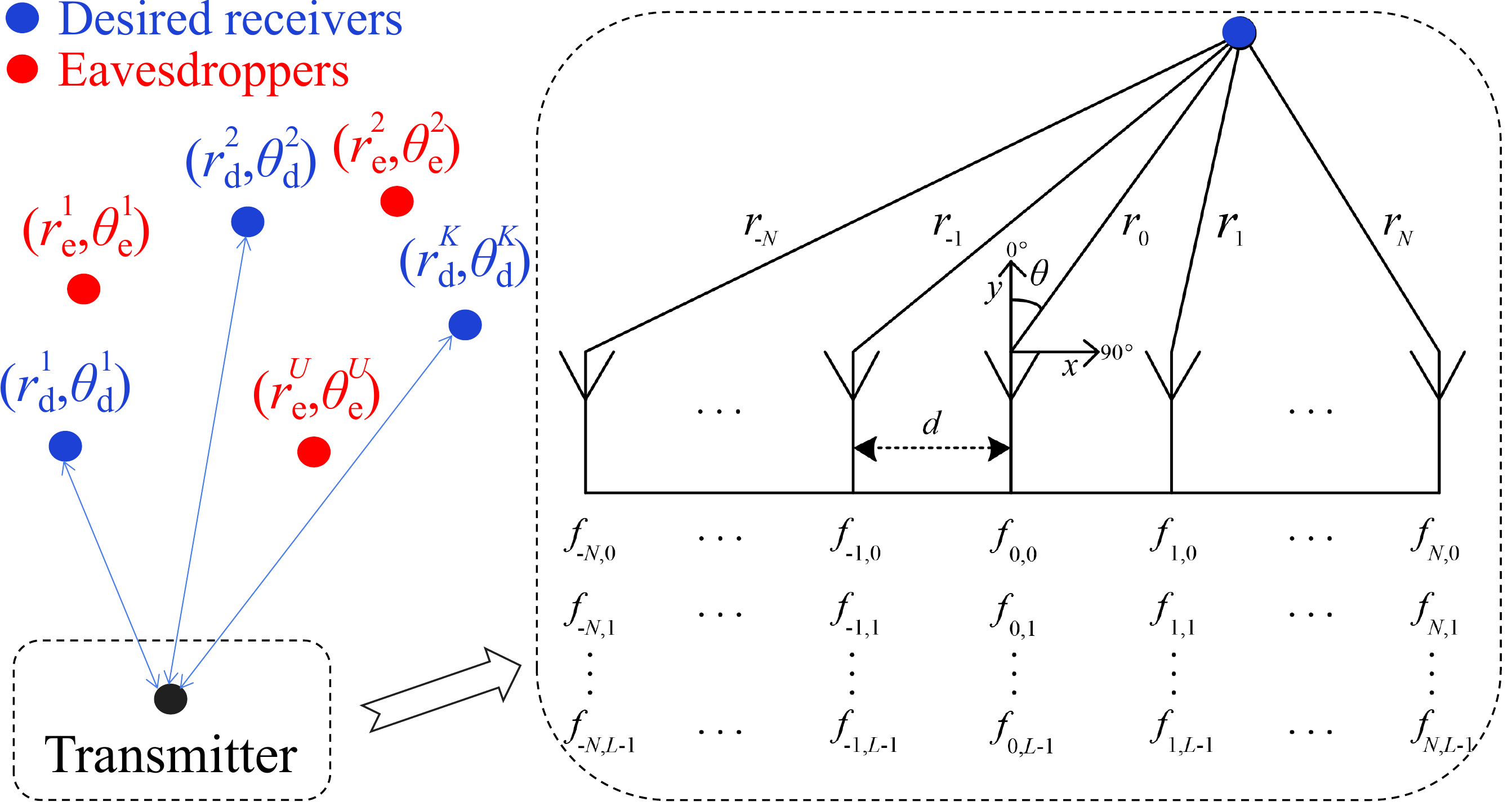}
\caption{System Model of multi-beam DM based on multi-carrier FDA.}
\end{figure}

\emph{Notations}: The operators ${( \cdot )^{\rm{T}}}$, ${( \cdot )^{\rm{H}}}$, ${( \cdot )^{-1}}$, and ${( \cdot )^{\dagger}}$ represent the transpose, Hermitian transpose, inverse, and Moore-Penrose inverse operations of a matrix, respectively.


\section{System Model}

As shown in Fig. 1, we consider a ($2N+1$)-element symmetrical linear FDA with multiple carriers \cite{Shao_Dot_FDA}. The spacing $d$ between adjacent elements is designed as half central wavelength, and $L$ carriers are transmitted via each element. The frequency of the $l$-th ($l=0,1,\cdots,L-1$) carrier for the $n$-th ($n=-N,\cdots,0,\cdots,N$) element of FDA is designed as
\begin{equation}
\label{eq_f_n_l}
f_{n,l}=f_{0}+\Delta f_{n,l}=f_{0}+\Delta f\ln \left[(|n|+1)(l+1)\right]
\end{equation}
where $f_{0}$ denotes the central carrier frequency, and $\Delta f$ is a fixed frequency offset satisfying $|\Delta f|\ll f_{0}$.

Let $r$ and $\theta$ represent the range between an arbitrary spatial position and the central FDA element, and the azimuth angle, respectively. The normalized steering vector radiated from the transmit antenna array at $(r,\theta)$ is a $(2N+1)L\times 1$ vector for such a system, which can be expressed as
\begin{equation}
\label{eq_h_r_theta}
\begin{aligned}
&{\mathbf{h}}(r,\theta)\\
&=\frac {1}{\sqrt{(2N+1)L}} \left [{\mathbf{a}}_{-N}^{\rm{T}}(r,\theta) \cdots {\mathbf{a}}_{n}^{\rm{T}}(r,\theta) \cdots {\mathbf{a}}_{N}^{\rm{T}}(r,\theta) \right]^{\rm{T}}
\end{aligned}
\end{equation} 
where ${\mathbf{a}}_{n}(r,\theta)=
\left[ 
e^{j\phi_{n,0}}  \cdots  e^{j\phi_{n,l}} \cdots  e^{j\phi_{n,L-1}}
\right ]^{\rm{T}}$ is an $L \times 1$ sub-steering vector caused by the $L$ carriers of the $n$-th antenna element \cite{Shao_Dot_FDA}. The phase of the $l$-th entry of ${\mathbf{a}}_{n}(r,\theta)$ is 
\begin{equation}
\label{eq_phi_n_l}
 \phi_{n,l}=2\pi\left[ \Delta f\ln[(|n|+1)(l+1)]\left (t-\frac{r}{c}\right)+\frac{ f_{0}nd\sin \theta}{c}\right] 
\end{equation}
where $c$ is light speed, $d=\left.\lambda \middle/ 2 \right.$ denotes the spacing between adjacent elements, $t$ refers to the time variable, and $\lambda=\left.c \middle/ f_{0} \right.$ represents the wavelength of the central carrier.

The multi-beam DM model consists of a transmitter with a $(2N+1)$-element FDA, $K$ stationary desired receivers, and $U$ passive eavesdroppers in different locations. The spatial coordinate of the $k$-th ($k=1,2,\cdots,K$) desired receiver is assumed to be $(r_d^k,\theta_d^k)$ and the combined set of the $K$ desired receivers' spatial coordinates is expressed as
\begin{equation}
\label{eq_R_Theta}
(\Upsilon_d,\Theta_d)=\left\{(r_d^{1},\theta_d^{1}),(r_d^{2},\theta_d^{2}),\cdots,(r_d^{K},\theta_d^{K})\right\}
\end{equation} 


A steering matrix $\mathbf{H}(\Upsilon_d,\Theta_d)$ with size $(2N+1)L\times K$ can be obtained by combining the steering vectors $\mathbf{h}(r_d^k,\theta_d^k)$ of the $K$ desired spatial positions $(\Upsilon_d,\Theta_d)$, i.e.,
\begin{equation}
\label{eq_H_R_Theta}
\mathbf{H}(\Upsilon_d,\Theta_d)=\left[\mathbf{h}(r_d^{1},\theta_d^{1})~\mathbf{h}(r_d^{2},\theta_d^{2})~\cdots~\mathbf{h}(r_d^{K},\theta_d^{K})\right]
\end{equation} 

Let $\mathbf{x}_d=\left[x_d^1~x_d^2~\cdots~x_d^K\right]^{\rm{T}}$ denote the confidential baseband symbol vector for the $K$ desired receivers. The weighted $N$-tuple transmitted signal vector of the FDA is designed as
\begin{equation}
\label{eq_s}
\mathbf{s}=\beta_1\sqrt{P_s}\mathbf{P}_{1}\mathbf{x}_d+\alpha\beta_2\sqrt{P_s}\mathbf{P}_{2}\mathbf{z}
\end{equation} 
where $P_s$ is the total transmit power constraint, $\beta_1$ and $\beta_2$ are the power splitting factors satisfying $\beta_1^2+\beta_2^2=1$, $\alpha$ denotes the normalization factor for the inserted AN, and $\mathbf{z}$ is the inserted complex AN vector with each entry having zero mean and variance $\sigma_z^2$. In order to obtain a desired standard modulation constellation at the desired receivers while distorting the received signals at other undesired receivers, the normalization matrix $\mathbf{P}_1$ and the orthogonal matrix $\mathbf{P}_2$ should satisfy the following criteria:
\begin{equation}
\label{eq_criteria}
\mathbf{H}^{\rm{H}}(\Upsilon_d,\Theta_d)\mathbf{P}_1=\mathbf{I},~~
\mathbf{H}^{\rm{H}}(\Upsilon_d,\Theta_d)\mathbf{P}_2=\mathbf{0}
\end{equation} 

In this paper, the normalization matrix is directly designed as the Moore-Penrose inverse matrix of the steering matrix $\mathbf{H}^{\rm{H}}(\Upsilon_d,\Theta_d)$, i.e.,
\begin{equation}
\label{eq_P1}
\begin{aligned}
\mathbf{P}_{1}=\mathbf{H}(\Upsilon_d,\Theta_d)\left[\mathbf{H}^{\rm{H}}(\Upsilon_d,\Theta_d)\mathbf{H}(\Upsilon_d,\Theta_d)\right]^{-1}
\end{aligned}
\end{equation} 
which satisfies $\mathbf{H}^{\rm{H}}(\Upsilon_d,\Theta_d)\mathbf{P}_{1}=\mathbf{I}_{K}$, and the size of which is $(2N+1)L\times K$.

It is known that the null space of a matrix can be derived from its SVD. Therefore, in order to design the orthogonal matrix $\mathbf{P}_{2}$, we perform the SVD operation on the steering matrix $\mathbf{H}^{\rm{H}}(\Upsilon_d,\Theta_d)$, which is now expressed as
\begin{equation}
\label{eq_SVD}
\mathbf{H}^{\rm{H}}(\Upsilon_d,\Theta_d)=\mathbf{U}\left[\mathbf{D}~\mathbf{0}\right]\left[\mathbf{V}_{1}~\mathbf{V}_{0}\right]^{\rm{H}}
\end{equation} 

Next, we can obtain a null space, that is, $\mathbf{V}_{0}\in \mathbb{C}^{(2N+1)L\times (2N+1-K)}$ for $\mathbf{H}^{\rm{H}}(\Upsilon_d,\Theta_d)$ from (\ref{eq_SVD}), which means $\mathbf{H}^{\rm{H}}(\Upsilon_d,\Theta_d)\mathbf{V}_{0}=\mathbf{0}_{K\times (2N+1-K)}$. Therefore, the orthogonal matrix of the SVD method can be designed as
\begin{equation}
\label{eq_P2}
\mathbf{P}_{2}=\mathbf{V}_{0}
\end{equation}

\begin{table*}[!t]
\renewcommand{\arraystretch}{1.5}
\caption{Complexity Comparisons Between ZF- and SVD-Aided Multi-Beam DM Systems}
\label{table1}
\centering
\begin{tabular}{ccc}
\hline
{\bf Items} &{\bf ZF method} &{\bf SVD method}\\ [0.1ex]
\hline
Orthogonal matrix $\mathbf{P}_2$ & $\mathbf{P}_2^{\rm{ZF}}=\mathbf{I}_{(2N+1)L} - \left[\mathbf{H}^{\rm{H}}(\Upsilon_d,\Theta_d)\right]^{\dagger}\mathbf{H}^{\rm{H}}(\Upsilon_d,\Theta_d)$ & $\mathbf{H}^{\rm{H}}(\Upsilon_d,\Theta_d)=\mathbf{U}\left[\mathbf{D}~\mathbf{0}\right]\left[\mathbf{V}_{1}~\mathbf{V}_{0}\right]^{\rm{H}}$,~$\mathbf{P}_2^{\rm{SVD}}=\mathbf{V}_0$\\
Size of $\mathbf{P}_2$ & $(2N+1)L\times (2N+1)L$ & $(2N+1)L\times (2N+1-K)$\\
Artificial noise $\mathbf{z}$ & $\mathbf{z}^{\rm{ZF}}\in \mathbb{C}^{(2N+1)L\times 1}$ &  $\mathbf{z}^{\rm{SVD}}\in \mathbb{C}^{(2N+1-K)\times 1}$\\
Time complexity of calculating $\mathbf{P}_2$ & ${\mathcal{O}}\left((2N+1)^{2}L^{2}K\right)$ &  ${\mathcal{O}}\left((2N+1)LK^2\right)$\\
Space complexity of storing $\mathbf{P}_2$ and $\mathbf{z}$ & ${\mathcal{O}}\left((2N+1)^{2}L^{2}\right)$ &  ${\mathcal{O}}\left((2N+1)L(2N+1-K)\right)$\\
\hline
\end{tabular}
\\
\end{table*}

The normalized line-of-sight (LoS) channel is considered in this paper. In fact, the proposed method can also hold over fading channels like  Nakagami-\emph{m} fading \cite{Ji_FDA_DM_Fading}, as long as the channel state information (CSI) is perfectly estimated. After passing through an LoS channel, the combined vector of the received signals at the desired receivers can be expressed as
\begin{equation}
\label{eq_y_d}
\begin{aligned}
\mathbf{y}(\Upsilon_d,\Theta_d)&=\mathbf{H}^{\rm{H}}(\Upsilon_d,\Theta_d)\mathbf{s}+\mathbf{w}_{d}
= \underbrace {\beta_{1}\sqrt{P_s}\mathbf{H}^{\rm{H}}(\Upsilon_d,\Theta_d)\mathbf{P}_{1}\mathbf{x}_{d}}_{\rm{Useful~Signal}}\\
&~~~+\underbrace {\alpha\beta_{2}\sqrt{P_s}\mathbf{H}^{\rm{H}}(\Upsilon_d,\Theta_d)\mathbf{P}_{2}\mathbf{z}}_{\rm{Artificial~Noise}}+\underbrace {\mathbf{w}_{d}}_{\rm{AWGN}}\\
&=\beta_{1}\sqrt{P_s}\mathbf{x}_{d}+\mathbf{w}_{d}
\end{aligned}
\end{equation} 
where $\mathbf{w}_{d}\sim {\cal CN}(\mathbf{0}_{K\times 1},{\sigma_{w_{d}} ^2}\mathbf{I}_{K})$ is the circularly symmetric complex additive white Gaussian noise (AWGN) vector with each entry having zero mean and variance $\sigma_{w_{d}} ^2$.

Similarly, the received signal of an arbitrary eavesdropper at $(r_e,\theta_e)$, $(r_e,\theta_e)\neq (r_d^k,\theta_d^k)$, can be expressed as
\begin{equation}
\label{eq_y_e}
\begin{aligned}
y(r_e,\theta_e)&=\mathbf{h}^{\rm{H}}(r_e,\theta_e)\mathbf{s}+{w}_{e} 
= \underbrace {\beta_{1}\sqrt{P_s}\mathbf{h}^{\rm{H}}(r_e,\theta_e)\mathbf{P}_{1}\mathbf{x}_{d}}_{\rm{Distortion~Signal}}\\
&~~~+\underbrace {\alpha\beta_{2}\sqrt{P_s}\mathbf{h}^{\rm{H}}(r_e,\theta_e)\mathbf{P}_{2}\mathbf{z}}_{\rm{Artificial~Noise}}+\underbrace {{w}_{e}}_{\rm{AWGN}}
\end{aligned}
\end{equation} 
where ${w}_{e}\sim {\cal CN}(0,{\sigma_{w_{e}} ^2})$  is the AWGN with zero mean and variance ${\sigma_{w_{e}} ^2}$. It is worth emphasizing that the first term of (\ref{eq_y_e}) denotes distortion signal for the undesired receiver, and the second term is the inserted AN which cannot be eliminated at the undesired receiver due to the fact that the designed $\mathbf{P}_2$ is non-orthogonal to its steering vector $\mathbf{h}^{\rm{H}}(r_e,\theta_e)$.

\section{Performance Analysis}


\subsection{Secrecy Rate}

We assume the $U$ eavesdroppers are located at  $(r_e^u,\theta_e^u)$, $1\le u\le U$, respectively, which  satisfy  $(r_e^u,\theta_e^u)\neq (r_d^k,\theta_d^k)$ for $\forall u\in\left\{1,2,\cdots,U\right\}$ and $\forall k\in\left\{1,2,\cdots,K\right\}$.

The achievable rate of the link from the transmitter to the $k$-th desired receiver can be calculated by \cite{Ji_FDA_DM_Fading}
\begin{equation}
\label{eq_R_d}
R(r_{d}^{k},\theta_{d}^{k})=\log_{2}\left[1+\gamma(r_{d}^{k},\theta_{d}^{k})\right]
\end{equation} 
where $\gamma(r_{d}^{k},\theta_{d}^{k})$ is the ratio of signal to interference plus noise (SINR) at the $k$-th desired receiver with the expression of 
\begin{equation}
\label{eq_SINR_d}
\begin{aligned}
\gamma(r_{d}^{k},\theta_{d}^{k})&=\frac{\beta_{1}^{2} P_{s} \mathbf{h}^{\rm{H}}(r_{d}^{k},\theta_{d}^{k}) \mathbf{P}_{1} \mathbf{P}_{1}^{\rm{H}} \mathbf{h}(r_{d}^{k},\theta_{d}^{k})} {\sigma_{w_d}^{2} + \alpha ^{2}\beta_{2}^{2} P_{s} \mathbf{h}^{\rm{H}}(r_{d}^{k},\theta_{d}^{k}) \mathbf{P}_{2} \mathbf{P}_{2}^{\rm{H}} \mathbf{h}(r_{d}^{k},\theta_{d}^{k})}
\end{aligned}
\end{equation}

Similarly, the achievable rate of the link from the transmitter to the $u$-th eavesdropper is expressed as
\begin{equation}
\label{eq_R_e}
R(r_{e}^{u},\theta_{e}^{u})=\log_{2}\left[1+\gamma(r_{e}^{u},\theta_{e}^{u})\right]
\end{equation} 
where
\begin{equation}
\label{eq_SINR_e}
\begin{aligned}
\gamma(r_{e}^{u},\theta_{e}^{u})&=\frac{\beta_{1}^{2} P_{s} \mathbf{h}^{\rm{H}}(r_{e}^{u},\theta_{e}^{u}) \mathbf{P}_{1} \mathbf{P}_{1}^{\rm{H}} \mathbf{h}(r_{e}^{u},\theta_{e}^{u})} {\sigma_{w_e}^{2} + \alpha ^{2}\beta_{2}^{2} P_{s} \mathbf{h}^{\rm{H}}(r_{e}^{u},\theta_{e}^{u}) \mathbf{P}_{2} \mathbf{P}_{2}^{\rm{H}} \mathbf{h}(r_{e}^{u},\theta_{e}^{u})}
\end{aligned}
\end{equation}

Therefore, the secrecy rate of the proposed multi-beam DM system can be defined as
\begin{equation}
\label{eq_secrecy_rate}
R_s=\max\limits_{k\in\left\{1,\cdots,K \right\}}\left[\min\limits_{u\in\left\{1,\cdots,U \right\}}\left(R(r_{d}^{k},\theta_{d}^{k})-R(r_{e}^{u},\theta_{e}^{u})\right)  \right]^{+} 
\end{equation}
where $[\cdot]^{+}=\max\{0,\cdot\}$.

\subsection{Time Complexity}

Both the normalization matrices of the ZF method in \cite{Xie_AN_ZF}\cite{Xie_FDA_DM} and the proposed SVD method are designed as $\left[ \mathbf{H}^{\rm{H}}(\Upsilon_d,\Theta_d)\right]^{\dagger}$, so the complexity depends on the orthogonal matrix $\mathbf{P}_{2}$ and the inserted AN $\mathbf{z}$, as shown in Table I.

Here, we consider the time complexity of calculating the orthogonal matrix $\mathbf{P}_2$. For the ZF method,  the orthogonal matrix is designed as  $\mathbf{P}_2^{\rm{ZF}}=\mathbf{I}_{(2N+1)L} - \left[\mathbf{H}^{\rm{H}}(\Upsilon_d,\Theta_d)\right]^{\dagger}\mathbf{H}^{\rm{H}}(\Upsilon_d,\Theta_d)$, the time complexity of which is ${\mathcal{O}}\left((2N+1)^{2}L^{2}K\right)$ \cite{Xie_AN_ZF}\cite{Handbook_Algebra}. For the SVD method, the orthogonal matrix can be directly obtained from the SVD of $\mathbf{H}^{\rm{H}}(\Upsilon_d,\Theta_d)$, which means the time complexity of calculating $\mathbf{P}_2^{\rm{SVD}}$ lies in the time complexity of calculating the SVD of $\mathbf{H}^{\rm{H}}(\Upsilon_d,\Theta_d)$. Since the size of $\mathbf{H}^{\rm{H}}(\Upsilon_d,\Theta_d)$ is $K\times (2N+1)L$, ($K<(2N+1)L$), the time complexity of calculating $\mathbf{P}_2^{\rm{SVD}}$ can be obtained by ${\mathcal{O}}\left(K^2(2N+1)L\right)$ \cite{Handbook_Algebra}.

Additionally, the Moore-Penrose inverse operation of a matrix can be acquired using SVD algorithm \cite{Handbook_Algebra}, which means the normalization matrix $\mathbf{P}^{\rm{SVD}}_1$ can be directly calculated using the output of SVD. But for the ZF method, the normalization matrix $\mathbf{P}^{\rm{ZF}}_1$ has to be calculated independently.

\begin{figure*}
\centering
\includegraphics[angle=0,width=0.88\textwidth]{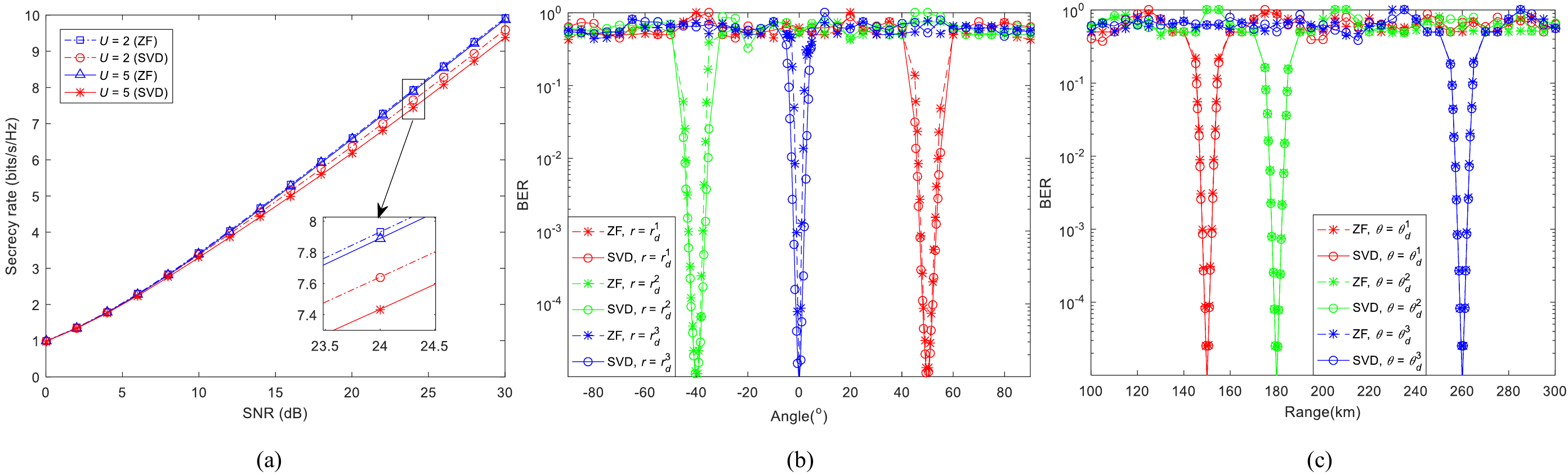}
\caption{Secrecy rate and BER performances of ZF- and SVD-aided DM systems. (a) Secrecy rate versus SNR; (b) BER versus angle; (c) BER versus range.}
\end{figure*}

\begin{figure*}
\centering
\includegraphics[angle=0,width=0.9\textwidth]{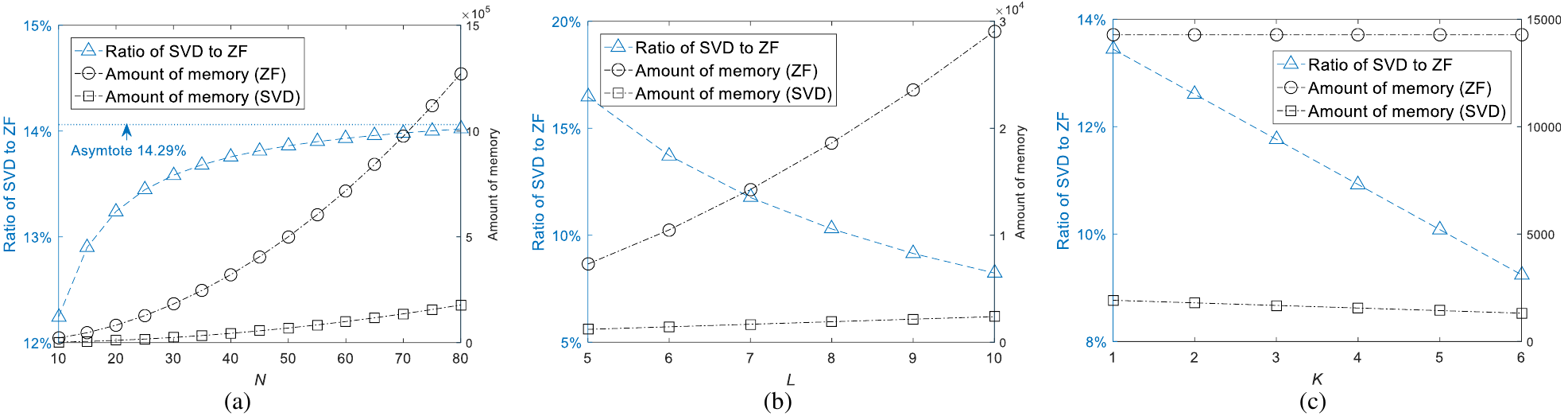}
\caption{Total amount of memory and ratio of SVD to ZF required by the orthogonal matrix $\mathbf{P}_{2}$ and the AN $\mathbf{z}$ versus (a) $N$ with $L=7$ and $K=3$; (b) $L$ with $N=8$ and $K=3$; (c) $K$ with $N=8$ and $L=7$.}
\end{figure*}

\subsection{Space Complexity}

The sizes of the orthogonal matrix $\mathbf{P}_{2}$ and the AN $\mathbf{z}$ can impact the the memory consumption significantly. As shown in Table I, the size of the orthogonal matrix $\mathbf{P}_{2}^{\rm{ZF}}$ of the ZF method is $(2N+1)L\times (2N+1)L$, and the size of the inserted AN $\mathbf{z}^{\rm{ZF}}$ is $(2N+1)L\times 1$. By contrast, the size of the orthogonal matrix $\mathbf{P}_{2}^{\rm{SVD}}$ of the proposed SVD method is $(2N+1)L\times (2N+1-K)$, and the size of the inserted AN $\mathbf{z}^{\rm{SVD}}$ is $(2N+1-K)\times 1$.

We define a metric $\zeta$ as the ratio of the total memory consumed by $\mathbf{P}_{2}^{\rm{SVD}}$ and $\mathbf{z}^{\rm{SVD}}$ to that of the ZF method, i.e.,
\begin{equation}
\label{eq_ratio}
\begin{aligned}
\zeta&=\frac{(2N+1)L\times (2N+1-K)+(2N+1-K)}{(2N+1)L\times (2N+1)L + (2N+1)L} \times 100\%\\
&=\frac{(2N+1)^{2}L- (2N+1)(LK-1)-K}{(2N+1)^{2}L^{2}+(2N+1)L} \times 100\%
\end{aligned}
\end{equation}

It is worth noting that $\zeta\to \left.1\middle/L\right.$ when $K$ is determined and $N\to\infty$, which means the proposed SVD method can consume at most $\left.1\middle/L\right.$ of the memory required by the ZF method, thereby reducing the amount of memory and lowering DC power consumption requirements.

\section{Simulation Results}

In our simulations, the parameters are set as $f_{0}=10$ GHz, $\Delta f=2$ kHz, $\beta_1=0.9$, $P_s=1$, $N=8$, $L=7$, $K=3$, and Gray-coded $\left.\pi\middle/4\right.$-QPSK, respectively. The signal-to-noise ratio (SNR) is set as $10$ dB. Without loss of generality, we further assume $\sigma_{w_d}^2=\sigma_{w_e}^2$, and the locations of the $K$ desired receivers are $(r_d^1,\theta_d^1)=(150\rm{km},50^{\circ})$, $(r_d^2,\theta_d^2)=(180\rm{km},-40^{\circ})$, and  $(r_d^3,\theta_d^3)=(260\rm{km},0^{\circ})$, respectively.

Fig. 2(a) shows the secrecy rate versus SNR (dB) for ZF- and SVD-aided multi-beam DM systems, where the eavesdroppers' locations are randomly selected in the simulations. It is observed that the SVD method requires slightly higher SNR (dB) than the ZF method in order to achieve the same secrecy rate. For example, only 0.5 dB of additional SNR is required for the SVD method to match the ZF method when $R_s = 8$ bits/s/Hz and $U=2$. The secrecy loss is due to that the size of the inserted AN of the SVD method is smaller than that of the ZF method, which makes the SINRs of eavesdroppers a little higher than that of the ZF method.

Fig. 2(b) and (c) illustrate the simulated BER performances for ZF and SVD methods versus angle and range, respectively. In the angle dimension, the BER lobes of the ZF method are slightly narrower than the proposed SVD method around the desired receivers. In the range dimension, both ZF and SVD methods can achieve almost the same BER performances.

In order to illustrate the time complexity, we conducted $10^4$ Monte Carlo experiments to record the average time consumption of calculating $\mathbf{P}_{2}$ using MATLAB{\protect\footnotemark[1]\footnotetext[1]{Computer configurations: Intel(R) Xeon(R) CPU E5-1620 v2 @ 3.7 GHz; 8.0 GB RAM; 64-bit operating system. MATLAB version: R2016a.}}. The result shows that the average time consumptions of ZF and SVD methods are $0.2258$ and $0.1933$ ms, respectively, which verifies that the computational efficiency of the proposed SVD method is better than that of the ZF method.

Moreover, Fig. 3 shows the total amount of memory required by the orthogonal matrix $\mathbf{P}_{2}$ and the AN $\mathbf{z}$, and the ratio $\zeta$ of SVD to ZF versus $N$, $L$ and $K$, respectively. From Fig. 3(a), the SVD method only consumes up to $14.29\%$ of the memory required by the ZF method with $L=7$ and $K=3$, which is due to $\zeta\to \left.1\middle/L\right.$ when $N\to \infty$ as shown in (\ref{eq_ratio}). From Fig. 3(b) and Fig. 2(c), the ratio of SVD to ZF decreases as $L$ and $K$ increases. Therefore, the SVD method is more efficient than the ZF method with respect to memory required.


\section{Conclusion}
A new SVD-aided low-complexity range-angle dependent multi-beam DM scheme based on symmetrical multi-carrier FDA was proposed. The proposed SVD method outperforms the ZF method in regard to the computational complexity and the amount of memory required for processing, while introducing only a small performance loss of secrecy rate. The SVD method opens a way to reduce the implementing complexity and lower DC power consumption requirements for range-angle dependent multi-beam DM systems.

\ifCLASSOPTIONcaptionsoff
  \newpage
\fi

\vfill

\end{document}